# Libraries' Metadata as Data in the Era of the Semantic Web

Modeling a Repository of Master Theses and PhD Dissertations for the Web of Data

## Manolis Peponakis

**Summary**: *This study argues that metadata of library catalogs can stand autonomously, providing valuable information detached from the resources they point to and, therefore, could be used as data in the context of the Semantic Web. We present an analysis of this perception followed by an implementation proposal for a Master's thesis and PhD dissertation repository. The analysis builds on the flexibility of the Resource Description Framework (RDF) and takes into account the Functional Requirements for Bibliographic Records (FRBR) and Functional Requirements for Authority Data (FRAD) in order to reveal the latent academic network by linking its entities to a meaningful and computationally processable set. Current library catalogs retrieve documents to find answers, whereas in our approach catalogs can provide answers that could not be found in any specific document.*

## *Introduction*

Library catalogs' metadata used to serve mainly as the mediator for reaching the data, namely the resources, and were disregarded as sources of information beyond that. The main reason for their existence was the allocation of the resource. This study argues that library catalogs' metadata could stand autonomously and provide valuable information regardless of the resources they point to. However, we need to reconsider certain issues about library catalogs -concerning, essentially, metadata- so that they are an interoperable part of the Semantic Web and not standalone databases. Since the rise of the Semantic Web most studies have been focusing on how we can transcribe existing data to the new "language" of the Semantic Web, namely the Resource Description Framework (RDF). Most of these efforts have been targeting the structural-syntactical aspects of the transition and hardly considered the new capabilities for semantic expressiveness which are provided by the new language.

In this new hybrid environment part of the Library and Information Science practices have gotten so lost in the syntax that they missed the semantics. Clearly "*To encode any knowledge in a formal representation requires the author of that knowledge to learn the representation's syntax and*

*semantics*"(Marshall & Shipman, 2003, p. 61). And yet, depending on the capabilities of the syntactical-structural level, the expression of semantics becomes more or less easy and accurate. Not long ago we experienced a similar situation. As soon as XML was introduced, experts were hands-on with the transition from ISO 2709 to MARC XML without considering the expressiveness of XML in order to further exploit its capabilities[1].

A key issue is always "*the separation of languages of description from the specific data formats into which they have for so long been embedded*" (Baker, 2012, p. 130). Towards this end Functional Requirements for Bibliographic Records (FRBR) was an innovative initiative because they tried -though not totally successfully- to do so. Unfortunately, IFLA's choice to adopt the Entity Relationship (ER) model, which entails a large number of syntactical restrictions, limited the expressiveness of the FRBR model. Still, this does not affect the novelty of the FRBR since its cataloguing theory brings important changes to the paradigm of cataloging. The most important contribution is the introduction of relations between different entities as the basis of description instead of the approach where the description of the resource, namely the record, was the one and only entity while various elements-fields served as its attributes. "*In this new reality, relationships can be recorded explicitly, allowing users to navigate between related resources, thus turning catalogs into true information networks, overcoming limitations of the current catalogs conceived basically as a set of bibliographic records*" (Picco & Ortiz-Repiso, 2012, p. 625).

At this point we need to stress the fact that the main subject of this study is neither the implementation of FRBR nor its successor, namely the Resource Description and Access (RDA), through RDF. There is extensive bibliography on this topic illustrating its assets and many researchers have proposed various approaches (Agenjo, Hernández, & Viedma, 2012; Alemu, Stevens, Ross, & Chandler, 2012; Dunsire, Hillmann, & Phipps, 2012; Dunsire, 2012; Hillmann, Coyle, Phipps, & Dunsire, 2010; Howarth, 2012; Takhirov, Duchateau, & Aalberg, 2011; Taniguchi, 2013; Yee, 2009). To our belief the most important contribution to this direction is the re-expression of the FR model with an object-oriented approach in the context of the harmonization with CIDOC-CRM which resulted to FRBR$_{OO}$ (Bekiari, Doerr, & Le Bœuf, 2012; Doerr & LeBoeuf, 2007).

Our work focuses on the possibilities offered by the new technological infrastructure in order to enrich the semantics and the expressiveness of catalogs' metadata, having always in mind that RDF does not indicate what to say but allows expressing whatever we want to say with specific rules and restrictions. This article does not argue about structural issues -at least to the point that structure does not affect expressiveness- such as whether RDF has been expressed in RDF/XML, N-triples or Turtle. It does argue, though, for RDF capabilities which offer the option of more coherent and sophisticated expression in a variety of domains. Besides, both RDF and "*Linked Data is a content agnostic technology, in that it can be used to publish all types of information: sociology, journalism, physics, philosophy, art, etc.*" (Summers, 2013). Additionally, the "*introduction of the semantic web standards have provided new integration methods, which are not only applicable within digital repositories' records, but*

---

[1] An exception to this rule is the transformation of UNIMARC's Contents Note (field 327). According to update 3, this field was restructured in order to carry the hierarchical tree structure of the Extensible Markup Language (XML) in ISO 2709 format.

*also could be utilized for integration of the records and the other types of web resources*" (Hakimjavadi & Masrek, 2012, p. 58).

The remainder of the study is organized as follows. The first part argues the potential of seeing the catalogs as the deconstruction of the resources' concepts into entities and linking these entities in various ways. The second part of the study proposes the implementation of this approach in terms of a Master theses and PhD dissertations' repository. Taking into account RDF in conjunction with FRBR and the Functional Requirements for Authority Data (FRAD) we recommend an implementation of the approach described previously in order to reveal the relations of the academic network by linking its entities, like professors, universities and dissertations, to a meaningful and computationally processable set.

## *Libraries' metadata as Linked Data*

In this section we argue that the metadata of library catalogs[2] should quit treating the resource as an indivisible object that is the center of the description. The resource description should be perceived as a set of structured entities containing small pieces of information, which do not necessarily derive directly from a single resource. To clarify this let us consider that it is likely that none of the Manifestations of Hamlet a library possesses mention the historical context or the time period in which Shakespeare wrote this play; but it is possible that the *Work* record contains this information. It becomes obvious that the resources could not substitute the catalog's information as the catalog is not just a sophisticated tool to find items/resources. To further support this we build upon two basic principles of the current trends in Information Technology and Library & Information Studies. The first one is the RDF and the second one is the FR family, especially FRBR and FRAD. The aim is to reveal the multiple dimensions of library catalogs' metadata.

There are two basic aspects related to the debate between the two approaches, meaning the ones which support one record per resource -such as Dublin Core (DC)- and the ones favoring RDF[3]. The first approach argues that a record contains all the necessary information about the resource. The second argues that a resource can be described using a set of autonomous statements each one of them being true (Baker, 2012). In the first case it is considered that the fields of a record can have no meaning out of this context. In the second case it is argued that every single statement can carry a definite meaning. The latter allows for a remarkable flexibility and makes statements easy to reuse. "*The concept of a resource is generalized in RDF to mean anything that can be described with metadata. This allows metadata to be applied to anything that can be identified, even if it cannot be directly retrieved on the web*" (Hillmann et al., 2010). Besides, "*From the perspective of a librarian, cataloger, publisher, or*

---

[2] In this work the coverage of the term "library catalog" extends from traditional OPACs to digital libraries and repositories.

[3] The truth is that most libraries of recent times do not fully abide by this logic since authorities are individual records. This applies to both MARC and newer approaches such as the Metadata Object Description Schema (MODS) and the Metadata Authority Description Schema (MADS). Even in these cases the rule of one-record-per-resource still applies because, in essence, the authority was incorporated to the record as a heading. Repositories seem to bring some unfortunate changes to this practice.

*content provider, the Semantic Web is a metadata initiative*" (Marshall & Shipman, 2003, p. 62). The motivation of our article is this very need to explore the new means of expression to its full potential and let go of ankyloses of older, record-based practices.

Notwithstanding the fact that the debate of object oriented cataloging versus the Entity Relationship (ER) model is known and quite dated (Heaney, 1995), FRBR was built using the ER model. There are different perceptions of things in these two approaches and the choice of the ER model seems to lead to many implementation problems. In the following, we argue through an example one of the fundamental differences of the two models.

According to FRBR and FRAD *Place* is both an entity as well as an attribute of several entities. Thus, *Place* exists in "Place of publication/distribution" as an attribute of a *Manifestation*, "Place of birth, Place of death, Place of residence" as attributes of a *Person*, "Places associated with family" as an attribute of a *Family*, "Place associated with the corporate body" as an attribute of a *Corporate Body*, "Term for the Place" as an attribute of a *Place,* "Place of origin of the work" as an attribute of a *Work* (IFLA, 1998, 2009). If we as well consider the "Location of agency", we come down to nine attributes that may refer to the same place, for example a city. Due to the fact that the ER model does not allow links between attributes, there is no way to indicate the semantic uniqueness of a value using the ER attributes. One way to avoid this problem is by examining the value of each attribute. If the values are the same, we could assume that the place referred to is the same. However, this practice could lead to the assumption that Athens stands for a single place altogether while Athens (Georgia) and Athens (Greece) are not the same. It is important to bring to the reader's attention the fact that appellation is not quite an accurate criterion to distinguish entities (Doerr, Riva, & Žumer, 2012).

Now let us assume that we have a way to address this specific place as a concept (regardless of its name) and indicate it using an identifier. This would allow for a single entry (on conceptual and not on literal level) of this place. Any other entity, like persons and events, which is somehow connected to this place, would be assigned the appropriate connection. This approach would treat the great majority of catalog elements as authorities. If the identifier used is in the form of a Uniform Resource Identifier (URI) based on http, then we have made a big step closer towards Linked Data and the Semantic Web. In this way, the aforementioned "one and only place" acquires its essence and can be used not as a part of an isolated database but universally.

Baker argues that "*RDF was designed for making statements about reality; DCAM was designed for specifying the contents of metadata records*" (Baker, 2012, p. 121). In this study we suggest the transformation of the statements about reality into metadata and vice versa. Moreover the "*term* [metadata] *is often used interchangeably with what is often more correctly called data, since all information is ultimately about something else, which is conceivable as information*" (Summers, 2013). In this manner the catalog becomes a kind of graph (Murray & Tillett, 2011; Peponakis, 2012) and its networked structure transforms it from lists of flat metadata records to knowledge structures and semantic networks.

But how is it possible to go down this road while the FR family does not point in this direction? Despite the unfortunate selection of the Entity Relationship model and the absence of "*generalization–specialization relations between classes and properties*" (Doerr et al., 2012, p. 530), the FR family builds on a crucial conception: The catalogs are the outcome of defined relationships between entities (more or less autonomous) and not the sum of indivisible records that describe indivisible resources. Beyond this underlies the perception that we can apply formal concept analysis to deal with these entities and their relations. Thus it is possible to perceive catalogs as networks which connect in meaningful ways the entities which constitute them. And RDF, rather the RDF Schema (RDFS), is the current tool for implementing this representation.

The RDF triples approach does not seem so innovative and revolutionary if seeing, faithful to older perceptions, the record for *Manifestation* as the kernel and the elements-fields as attributes of it. The classical descriptive metadata, in which information is simply transcribed from the resource, is a common practice, and such cases are already treated sufficiently through the current cataloging theory paradigm. In addition, such cases have no difficulties of being incorporated into the RDF based context by transforming the triple "subject-predicate-object" to "resource-property-value". The true power of RDF is the ability to use the triple "resource-property-resource" where each resource could be a subject or object to a new triple. The disadvantages of current library catalogs are not related to the formalism of values but to the formalism of connections between catalogs' entities (both within a single catalog and among various catalogs). Formalizing the inter and intra connections could add new relations and also reduce the identical information which, according to some researchers, is "*'polluting' search engine results with massive amounts of redundant information*" (Gradmann, 2005, p. 64). On the other hand, a considerable amount of the catalogs' information does not belong to the category of descriptive metadata so it has no reason to be repeated as value. From the aforementioned nine occurrences of the FR's attribute *Place* only one, according to RDA, -namely the *"place of publication/distribution" as an attribute of a Manifestation-* must be transcribed from the specific resource and, thus, provide a specific literal taken from it.

We need to acknowledge that in the digital environment it is easy to move from one resource to another, for which a URI is adequate. Consequently, the user can examine the resource herself a lot more easily than within the environment of a traditional library. Therefore, to only create metadata containing much descriptive information is not so apt. It is more important to create metadata with meaningful and computational information rather than with a plethora of information that FRBR defines as inherent, i.e. attributes that "*can usually be determined by examining the entity itself*" (IFLA, 1998, p. 31). According to the new paradigm, the starting point of cataloging must be seen as "*a process of making observations on resources*" (Murray & Tillett, 2011, p. 171) and not as a process of simply transcribing information from resources. If we further consider that the subject of research is to put these observations into a computational framework, this leads to the conclusion that we live in a period of profound change for libraries' catalogs. The true benefit of perceiving RDF statements as the basis of communication is the potential to link using a URI instead of downloading records, so that "*Linked Open Data is sharable, extensible, and easily re-usable*" (Baker et al., 2011). It is, then, easy to build on information created by others, in a way that, instead of referring to something using an appellation, one can use a URI to do so.

Let us consider a specific example. If someone wants to refer to the Greek novelist Nikos Kazantzakis she can use the URI of the Virtual International Authority File (VIAF) http://viaf.org/viaf/14771803. And if she wants to associate Kazantzakis to his place of birth, she -instead of just naming the island of Crete- can use the URI of the Library of Congress http://lccn.loc.gov/n79150324 which describes this entity. It is also possible to connect his work "The Last Temptation of Christ" with the IMDB's URI http://www.imdb.com/title/tt0095497/ and indicate that the screenplay of the movie was based on his novel. This way a network which is based on the relations between autonomous entities is deployed. As illustrated in the example, these relations and entities do not necessarily come from one specific domain. In this context the inspired title "From Collections to Connections" of the 2013 BOBCATSSS conference reveals one of the major transitions of our times in resource description and discovery.

To summarize, by using a URI instead of a name it is possible to allocate the information which has been assigned to the corresponding entity. This is an easy way to refer uniquely to this entity. The major challenge then is about the responsibility of the integrity concerning the entity's description. How accurate, valid and trustable is the description? For example, is it safe to use the URI, http://scholar.google.com/citations?user=4rO87mQAAAAJ to refer to Aristotle's citations? According to this source Aristotle is an "Ancient Greek philosopher" with "Verified email at buffalo.edu". Can we then assume that Aristotle has recently moved to New York and teaches at the University at Buffalo? Besides the comical aspect of this observation, which obviously lacks in truth, it is very important to check the provenance of the information. It is quite different to mention that blog X argues that the moon is a spacecraft than to refer to the official NASA's statements.

In the new environment, given that anyone can publish information, provenance becomes more and more important so that a very crucial, two-dimensional issue comes up: First it is the responsibility for the creation of specific data to which one can link and; second, it is the provenance verification of the information creation. Libraries deal with both aspects. They have both the know-how to create structured data and they are trusted information providers for others to re-use their data. Therefore, libraries ought to reconsider their role as publishers of structured information for the Semantic Web.

The next section discusses an implementation of the aforementioned approach in terms of a Master theses and PhD dissertations' repository and analyzes the role of the academic libraries in this context.

### _Modeling an ETDs' repository as a semantic network_

As Ivanovic et al state "_One of the basic postulates of a knowledge society is availability of knowledge_" (Ivanovic, Ivanovic, & Surla, 2012, p. 548). Overlooking the fact that postgraduate theses and dissertations are an important capital of the academic knowledge base (Fox, McMillan, & Srinivasan, 2009) libraries, in many cases, do not treat them accordingly. The term "grey literature" is used to describe this kind of content and signifies its separation from the catalog's mainstream material. In this section we propose a method that can transform grey literature to "bright white".

Since the early 2000s institutional repositories hosting Electronic Theses and Dissertations (ETDs) "*are gradually increasing in number and quality at higher-education institutions*" (Hakimjavadi & Masrek, 2012, p. 58). Current approaches emphasize flat metadata schemas to describe their resources (Park & Tosaka, 2010). The approach of the "Metadata Standard for Electronic Theses and Dissertations" (Atkins, Fox, France, & Suleman, n.d.) by the Networked Digital Library of Theses and Dissertations (NDLTD) is applied in the same context. A brief description of the metadata elements used in various schemas for theses and dissertations (such as DC, EDT-MS format, CERIF) can be found in (Ivanovic et al., 2012). Despite the fact that the problem of cataloging theses in order to retrieve the involved parties goes back in time (Harris & Huffman, 1985), little progress has been made so far. Most efforts aim at the dissemination of these documents on the Web through the web search engines. In this case the prerequisite is the implementation of an Open Archive Initiative Protocol for Metadata Harvesting (OAI-PMH), and in most cases this is considered sufficient.

In order to reduce the cost of cataloging and increase the value and reuse of the catalogs' data, this study proposes a slightly different orientation from the existing cataloging practice, which introduces major changes to the use of data. Schöpfel names metadata, interoperability, and services as three of five ways to add value to PhD dissertations; the other two being the quality of content and the format (Schöpfel, 2013). Here we discuss the former three under the prism of how metadata could be interoperable in a broader context and serve as the basis for added-value services. In this context the point is not just to describe Master theses and PhD dissertation as documents. It is more about the indication of interrelations between the entities that constitute these documents along with the depiction of their connections' network. As Johnson and Boock state "*the primary advantage of the linked data model over our Dublin Core and MARC records is the representation of people, academic departments, and degrees as independent resources. These concepts were previously represented either as flat text entries or, in the case of some names in MARC records, as name headings. They are now assigned URIs and can be the subject of metadata statements in their own right*" (Johnson & Boock, 2012, p. 4). The designers of the academic repositories must see their metadata from this point of view. In this way libraries will deal with the criticism of FRBR$_{OO}$ that "*FRBR$_{ER}$ envisions bibliographic entities as static, ever-existing things that come from nowhere*" (Bekiari et al., 2012, p. 12).

## Academic libraries as semantic information providers

Identifiers used to and still have a key role in libraries (Niu, 2013; Pisanski, Žumer, & Aalberg, 2010). Many systems have been developed from various institutions recommending ways for achieving unique identification in the digital environment, like the Digital Object Identifier (DOI), the "Handle system", the "AuthorClaim" and the Open Researcher and Contributor ID (ORCID) just to name a few. These efforts can be divided in two major categories depending on who is responsible for creating the record and managing the corresponding information. For example, the end user of ORCID has the main responsibility for her record. On the other hand in VIAF the user cannot obtain a URI herself. In the first case we rely on the claims of an individual, while in the second we rely on an institution. The difference is very important since provenance of information affects its trustability.

Going further into the academic context, it is very important for academic libraries to realize that they are responsible for the detailed and structured description of a large amount of their parent institutions' production. We claim that, in many cases, the analytic description that hides behind the http-based identifier must be provided by the institution through its library. There are three basic reasons for this. First the academic/research library knows how to encode information and create structured data. Second the library has access to the required information as the entities in question are either members or products of the institution. Finally, it is more likely for external users to trust an institution's claims than an individual's.

It is important that the authority file represents accurately the structure of the university or the research center. Libraries have both the tools and the know-how for formally representing the institution's structure such as the division of the university into schools, faculties and departments. Taking into account that this is not meant to be a static authority file it should also contain information on when a department was founded, when a change of name took place and so on. In this way we can have a formal representation of the university's structure created by those primarily responsible for its creation. This is trustworthy information upon which anyone can build.

The same applies for the members of the institution. Initiatives like *ORCID* and *AuthorClaim* can be useful but still the institution's library is the entity responsible for integrating the information. Besides, in many cases it is impractical or even impossible for the author to provide any kind of information, such as in the case of a deceased individual. The authors of this study are aware of the arguments about publishing personal data. However, the concern of this study is to show the potential of information management and not to make judgements on legal or moral issues about personal data. Emphasis should be given to the fact that even in argumentative situations, the involved library has in most cases the ability to directly consult with the individual whose data is required and take a direct statement of agreement for the matter in question[4].

Finally, the description of the documents is another issue of concern. In many cases authors supply the description of their documents but they may not be qualified to do so. To give an example, "*Although the students do not realize it they are creating the basis of the metadata records for their theses*" (Reeves, 2007). In addition, since we refer to the identification of relations between entities of the catalog and the encoded recording of this information, these are tasks for a trained expert who can create the basis of the knowledge upon which others will build. This way the information is coded appropriately and every entity involved in the production of the theses and dissertations of an institution has been described accurately and sufficiently. Furthermore, using the established vocabularies it is possible to publish this data on the Semantic Web. Knowing that this type of information is recorded by universities, it is -rather fairly- assumed that provenance ensures integrity. In other words, if integrity is ensured on account of the institution's signature, it is easier for others to accept, trust and reuse the metadata.

---

[4] We consider, as a representative example of the choice of publishing personal information, the gender transition of Professor Lynn Conway. See http://ai.eecs.umich.edu/people/conway/. Her choice was to come forward with her transition, while for others even the date of birth is considered private data.

**Expressing the semantics of an ETDs' repository basic entities**

With RDF we can make simple statements, analogous to natural language sentences. "*As in natural language, these URI-words fall into grammatical categories – classes and properties, roughly analogous to nouns and verbs.*" (Baker, 2012, p. 117). Using the semantics from FRBR and FRAD in conjunction with RDF we can describe quite accurately the academic network of theses and dissertations. As it is depicted in figures 1 and 2 we have *Works*[5], *Persons*, and *Corporate bodies*. The PhDs and Masters in figures 1 and 2 represent the *Work* entity of the FRBR. Taking into account that *Works* are "mutable, and dependent on reception for their interpretation" (Smiraglia, 2007, p. 182) for the purpose of this study it is considered that the *Work* is created during the time of the studying. This is because the written text of a thesis or dissertation comes from the studying procedure, so that the *Work's* "creation" takes place during this time period. This approach allows modelling the dissertation's synthesis as a process in order to look at the final *Item* as a result of this process. If we create a strict mapping to pure FRBR, then the property "*is student*" corresponds to "*is created by*", with inverse direction. Likewise the "*change to*" is the FRAD's "*Sequential relationship*", the "*has subdivision*" is FRAD's "*Hierarchical relationship*" and the "*is professor*" is a specialization of the FRAD's "*Membership relationship*"[6]. All other classes and relations/properties that appear in the graph are already defined, more or less clearly, from the MARC era. There are relator codes for the "*Degree-grantor*" (*295* in UNIMARC and *dgg* in MARC 21)", "*Thesis advisor*" (*727* in UNIMARC and *ths* in MARC 21) and "*Dissertant*" (*dis* in MARC 21)[7]. Finally, the gender of a *Person* is used transforming the FRAD attribute "*gender*" to a class.

Almost every triple is specialized by a time span as shown in Figure 1 and 2. The intention is to be clear that the linkage is valid for a specific time. The management of time through RDF is quite ambiguous and has sparked debate since the late 2000s. The basic reason for this is that there is not quite clear guidance on how to implement time. According to a W3C recommendation (available at http://www.w3.org/TR/2013/WD-rdf11-concepts-20130115/) "*The RDF data model is atemporal: It does not deal with time, and does not have a built-in notion of temporal validity of information.* [...] *However, RDF graphs can express information about events and about temporal aspects of other entities, given appropriate vocabulary terms*". To deal with this problem several ways have been proposed. Motik proposed the "temporal triple" which is defined as "*an expression of the form <s,p,o>[t] or <s,p,o>[$t_1$, $t_2$]*" (Motik, 2012, p. 7). A similar approach was introduced by Pugliese et al (Pugliese, Udrea, & Subrahmanian, 2008). The graphs appearing in our paper are based on these extensions. A different approach (without extensions on the basic RDF triple and thus absolutely compatible with RDF) was implemented by FRBR$_{OO}$ as it conformed to CIDOC-CRM[8]. In this case temporal entities were introduced. These temporal entities led to the addition of new (in relation to FRBR) classes (such as *Work Conception*). We find that the option to attach time on a triple allows for more flexibility in the case of ETDs. But as long as the community has not accepted it as a standard, its adoption would cause interoperability problems. Nevertheless, no matter how time could be modelled, we believe that the coding of it in order to be computationally processible is very essential.

---

[5] The *Work*, *Expression*, *Manifestation*, *Item* (WEMI) entities are not analyzed in the graphs both due to space limitations and because the relations between these entities are quite obvious in the FRBR model.

[6] Relations between different types of Authorities such as "*Membership relationship*" were a problem in MARC. FRAD took this a step further and introduced explicitly the relations between different types of Authorities. So according to FRAD, it is possible for a catalog to describe relationships between a family and its members, a corporate body and its members, and so on.

[7] The roles of advisor, supervisor their number and the number of viva (i.e. examining committee) members vary across universities worldwide.

Figure 1 shows the division of a University to Schools and Departments along with the assumed relationships between persons and theses as well as dissertations. Reading the statements of the graph presented in Figure 1 we (and a computer program as well) can assume that "Person A" is a man who earned a Master and a PhD at "Faculty E" and "Faculty B" respectively, which both are subdivisions of "University X". His master's thesis supervisor was "Person B" and his PhD supervisor was "Person C", both being female. We also get information about "Faculty B" which is a subdivision of "School A" and was established in 1963 and so on.

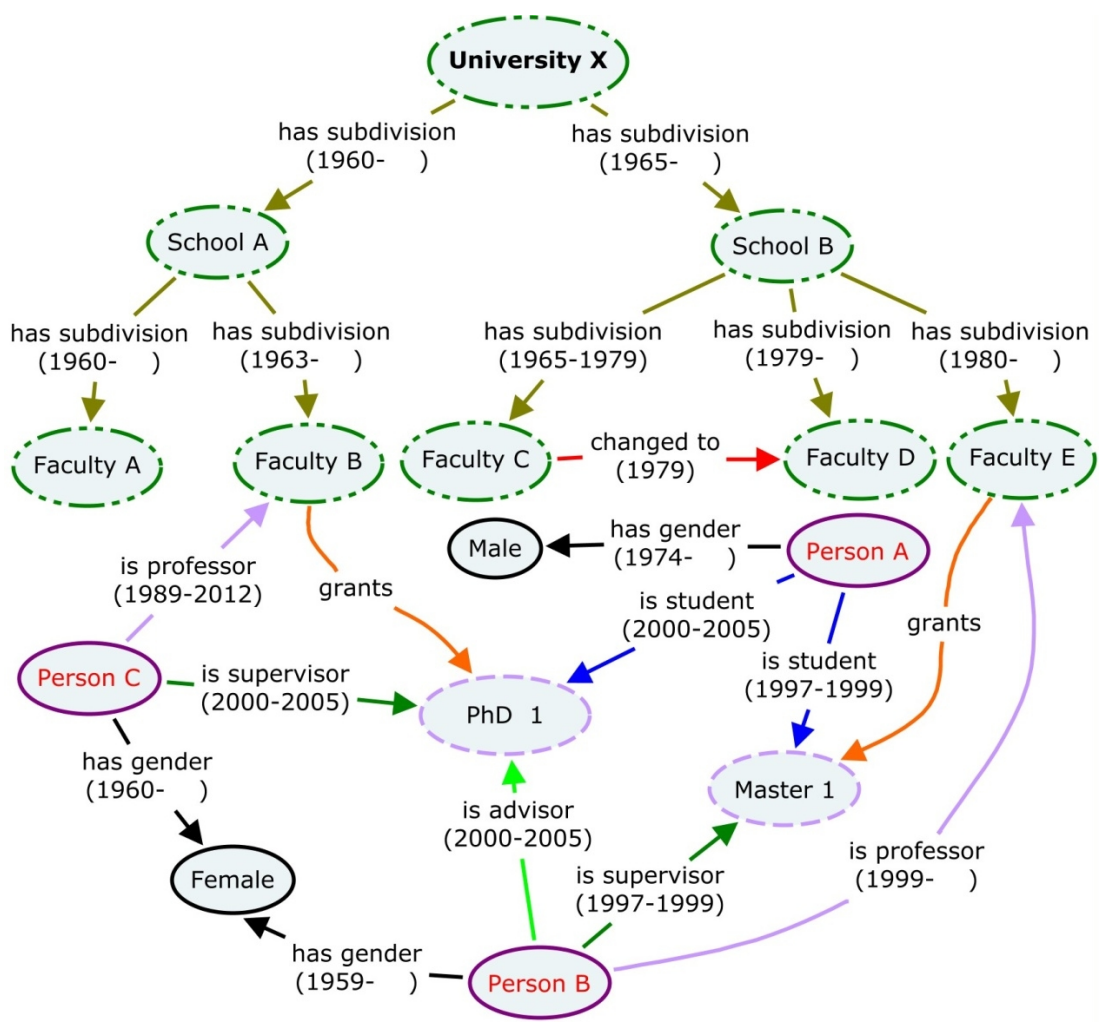

Figure 1: fragment of a visualization of a university's academic network

---

In Figure 2 the subdivisions of the universities (appearing in figure 1) are omitted due to lack of space and in order to give a more abstract view of the network. In addition to figure 1, this graph shows the persons' mobility between universities. According to figure 2, "Person A", six years after he was granted his PhD from "University X", became a Professor at "University Y", where he supervises a PhD candidate named "Person D". The PhD supervisor of "Person A" also moved to "University Y" 1 year later.

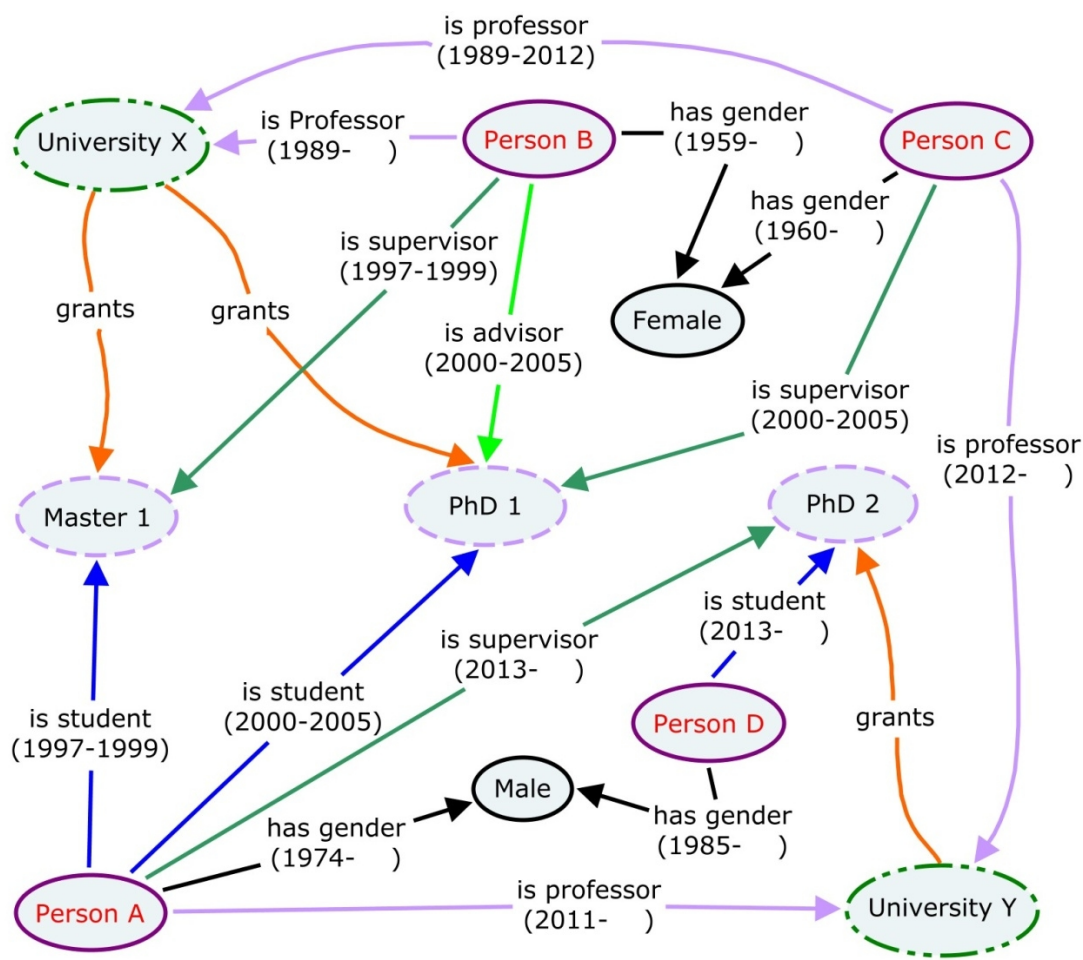

Figure 2: fragment of a visualization of an academic network in which the connections between universities appear

## Examples on the search and retrieval process

The dominant perception is that the word "search" is mostly used to indicate keyword searching, while the word "retrieval" is defined as the process of retrieving documents. Formal approaches of knowledge representation, though, offer a more sophisticated representation and allow not only the retrieval of relevant documents (i.e. records in the case of catalogs) but also several types of reasoning. The point is not to abandon keyword searching and the functions of catalogs as they operate currently but to take them one step ahead. Through the modeling described above we can answer several questions on a button's click. To clarify this we provide some examples based on the graphs of figures 1 and 2.

To begin with let us consider the structure of each institution along with types of information, such as when a department was established or split. We can count and locate each subdivision's faculty members and follow their mobility through the years in and out of the university as well as among the university's departments. We can also answer a question on the number of interdisciplinary postgraduate courses that have been launched through the years. The latter allows us to more precisely identify the intersected edges of the network, which is very important because, as Kwaśnik states, in some cases "*Overarching perspective [...]such as cultural studies cut across all school and college disciplinary boundaries making it difficult to neatly parse the university into clean academic categories*" (Kwaśnik, 2011, p. 12).

The aforementioned kind of network could be an important source of information particularly for the domain of gender studies. It allows gathering information on a number of issues, like the number of women and men in each faculty, the rate of women who supervise PhD students, and whether this number increases or decreases through time. Further, more questions can be answered on the number of men who are supervised by women and vice versa, on the number of women who are members of a viva (i.e. examining committee) in relation to the relative number of men, on the percentage of gender distribution among the departments and disciplines. Finally, through this kind of networks, we could even have the starting arguments concerning debates such as whether the domain of engineering is dominated by men and whether women's presence in humanities increases through the years.

If we introduce this modeling at a national or international level, we can create a wider network which provides answers to even more questions. Some of them could be about who taught where, what the mobility of professors according to their gender is, whether academic institutes cooperate with each other and at which level, or how the hierarchy is applied in universities' schools, departments, and so forth.

## *Discussion*

By and large libraries succeeded through various types of catalogs, such as OPACs and repositories, to manage the documents of their collections. Having now to manage the digital content too, they are standing at a crossroads, where the tangible objects of the past, such as books, separated their content from its carriers (Doerr & Tzitzikas, 2012). These liberated contents are being embodied in various resources and connected with multiple entities in a way that the creation of one record per resource seems a deficient simplification.

By modeling the catalog's metadata in the way we have described, the catalog will have been transformed to an information network, which emphasizes the relations and computational processes. In this context we consider FRBR "[not] *as simple metadata describing a specific resource but more like some kind of knowledge related to the resource*" (Peponakis, 2012, p. 599). As such, by using this knowledge, catalogs constitute an interoperable part of the Web's Linked Data. Besides, the innovation of Linked Data is that we can reuse existing data and extend them if and when required instead of continually reinventing the

wheel – which, in the scope of libraries, refers to the continual creating from scratch descriptions for things already described by others. Therefore, libraries must encode their data in order to achieve interoperability both among the libraries and with other content providers. It is important for this process to always bear in mind that an output could be used as input for further processing.

The challenge is not the keyword searching but the semantic querying and reasoning, and this affects the search and retrieval as we know it (Tzompanaki & Doerr, 2012). If we want "*pidgin for digital tourists*" (Baker, 2000), Dublin Core is sufficient enough. But if we want to practice deeper analysis, "pidgin" is not proper for scientific discourse. Thus, we must create formal approaches (currently expressed in terms of formal ontologies) that will allow different layers of abstraction. In this way, there is the potential to be more specific when we want to focus specifically on PhDs than when we deal with postgraduate studies in general and so on. At the top level of abstraction there will be a simple schema (such as DC), where the rich semantics of the original schemas are mapped to general elements by multiple simplifications. Consequently, these layers will allow our formalisms to extend between the axes of accurate complexities and rough simplifications ("pidgins") depending on specific needs. If our controlled vocabularies consist only of pidgin, it is very difficult to upgrade them to scientific discourse. On the other hand, if we have no layers, it is not easy to downgrade scientific discourse to something simpler in order to communicate with other domains.

For robust catalogs we must reconsider the theory and practice of flat metadata along with the one record per resource correspondence. The resource is not an indivisible object. In many cases it contains more than one *Work*; or it contains one *Work* but more than one *Expression*. FRBR allows this perception but, in order to take full advantage of its potential, its implementations (either involving original cataloguing or in FRBRization projects) should not remain faithful to old practices of resource-centric approaches. For an efficient description, the recourse must be deconstructed to its primary entities and the synthesis of these entities will allow a fruitful representation of the resource and its connections with other entities. From this perspective, catalogs do not just implement information retrieval techniques but they offer structured data that carry meaning.

"*The lessons of the past and the opportunities facing us combine to suggest strongly that continuing to interpret bibliographic control as a monolithic, top-down effort designed to achieve universality—as the library world has traditionally done—is not going to allow us to take advantage of new technologies or new ways of thinking about and building metadata. A new paradigm of bottom-up allows "control as co-ordination" through semantic mappings—making the essential shift from controlling the data to controlling the semantics that will allow us to move forward, taking our legacy data with us*" (Dunsire et al., 2012, p. 176). In order to get the data with us, we must reduce their dependence on specific software because the "*quality metadata are of lasting value*" (Miller, 2012), while specific software is doomed to live a very short life.

This study has mainly tried to reveal the latent network which hides in the catalogs. In this context it explored ways to bring to the surface the functional entities which constitute a resource, along with the identified intra-resource relations as well as the inter-resources relations. In this context, the libraries' catalogs carry data and not just metadata, so that a software agent or a human being could use them to find answers not just documents. In these days it would be unrealistic to claim that anyone would consider searching in a library's catalog to find where Science Fiction writers or poets live or to conduct research on the causal relationship of a birthplace to novel writing or to inclination for mathematics. Library catalogs which conformed to the semantics of FRBR have already coded this information. A slightly different point of view allows making it publicly available in a searchable and meaningful form. In this way current library catalogs that retrieve "books" to find answers will be transformed into catalogs that can provide answers that may not be found into any specific "book".